\documentclass[aps,prb,twocolumn,a4paper,10pt,showpacs]{revtex4-1}
\usepackage{amsmath}
\usepackage{amsfonts}
\usepackage{amssymb}
\usepackage{graphicx}
\usepackage{subfigure}
\usepackage[pdfborder={0 0 0}]{hyperref}  
\usepackage{color}
\usepackage[latin1]{inputenc}
\usepackage{float}
\newcommand{\ep}{\epsilon}
\begin{document}
\title{Anomalous scaling and super-roughness in the growth of CdTe polycrystalline films}
\author{Ang\'elica S. Mata}
\author{Silvio C. Ferreira Jr.}\email{silviojr@ufv.br}
\author{Igor R. B. Ribeiro}
\author{Sukarno O. Ferreira}
\affiliation{Departamento de F\'{\i}sica, Universidade Federal de Vi\c{c}osa, Vi\c{c}osa, Minas Gerais, 36570-000, Brazil}


\begin{abstract}
{CdTe films grown on glass substrates covered by fluorine doped tin oxide by Hot Wall Epitaxy (HWE) were studied through the interface dynamical scaling theory.} Direct measures of the dynamical exponent revealed an intrinsically anomalous scaling characterized by a global roughness exponent $\alpha$ distinct from the local one (the Hurst exponent $H$), previously reported [Ferreira \textit{et al}., Appl. Phys. Lett. \textbf{88}, 244103 (2006)]. A variety of scaling behaviors was obtained with varying substrate temperature. In particular, a transition from a intrinsically anomalous scaling regime with $H\ne\alpha<1$ at low temperatures to a super-rough regime with $H\ne\alpha>1$ at high temperatures was observed. The temperature is a growth parameter that controls both the interface roughness and dynamical scaling exponents. Nonlocal effects are pointed as the factors ruling the anomalous scaling behavior.
\end{abstract}

\maketitle

\section{Introduction}

The control of surface dynamics is a crucial step for production of thin film based optoelectronic devices since features as grain size and surface morphology are among the most important features affecting the efficiency of these devices.\cite{Contreras} Consequently, the surface of many thin films has been studied extensively in the last years \cite{Karabacak, Elsholz, Igor, Yim, Lafouresse, Igor2} using dynamical scaling exponents which, in theoretical studies, associate universality classes to distinct processes involved in the interface growth. \cite{Barabasi, Meakin} In particular, the dynamical scaling of CdTe films, one of the most promising materials for the production of high-efficiency solar-cells and other electronic devices,\cite{Aramoto,Rams} grown on amorphous substrates at varying temperatures has been characterized through the Hurst ($H$) and growth ($\beta$) exponents. \cite{Igor,Igor2,Leal} CdTe films grown on amorphous substrates exhibited a peculiar behavior: the growth exponent and the global interface width are increasing functions of the temperature. \cite{Igor,Igor2} 

The analysis restricted to $\beta$ and $H$ exponents may contain only part of the information about the dynamical scaling, since several models\cite{DasSarma,Castro,Lopez1,Ramasco,Lopez2,Bru} and experiments\cite{Yim,Lafouresse,Santa,Huo} have recently demonstrated the presence of anomalous scaling implying in a global roughness exponent distinct from the Hurst exponent, commonly called local roughness exponent. The interface evolution can be characterized by the interface width, defined as the root mean square deviation of the interface height around its mean value on a scale $\ep$ defined by
\begin{equation}
\label{eq:eq1}
w(\ep,t) = \left\langle \left( \overline{[h({x},t)- \overline{h} ]^2} \right)^{1/2} \right\rangle,
\end{equation} 
where the bar represents the average inside windows of size $\ep$ and $\langle \cdots \rangle$ the average over different profiles. The common behavior of the interface width of globally self-affine profile follows the Family-Vicsek scaling ansatz \cite{Lopez1,Family}
\begin{equation}
w(\ep,t) = t^{\beta} f\left(\frac{\ep}{\xi(t)}\right).
\end{equation}
The scaling function $f(u)$ is
\begin{equation}
\label{eq:scl}
f(u) \sim \left\lbrace 
\begin{matrix}
u^\alpha ~~~ & \mbox{~~if }u\ll 1\\ 
\mbox{const} & \mbox{~~if }u\gg 1
\end{matrix} 
\right.,
\end{equation}
where $\alpha$ is the roughness exponent. The horizontal correlation length grows with time as $\xi\sim t^{1/z}$ for sufficiently large substrates. In the literature, $z$ is called dynamical exponent and is related to the growth and roughness exponents by\cite{Barabasi} $\alpha=\beta z$. It is worth to stress that given the power laws $w\sim t^{\beta}$ and $\xi\sim t^{1/z}$  concomitantly with the scaling function (\ref{eq:scl}), the relation $\alpha = \beta z$ must be obeyed and, therefore, these exponents are not independent. 

However, the scaling behavior of the local interface fluctuations may differ from the global ones characterizing an anomalous scaling.\cite{Lopez1,Ramasco,Lopez2,Bru} Indeed, the interface width on a scale $\ep$ is given by
\begin{equation}
w(\ep,t)=t^\beta g\left(\frac{\ep}{\xi(t)}\right),
\end{equation}
where the anomalous scaling function is
\begin{equation}
g(u) \sim \left\lbrace 
\begin{matrix}
u^H ~~~ & \mbox{~~if }u\ll 1\\ 
\mbox{const} & \mbox{~~if }u\gg 1
\end{matrix} 
\right. .
\end{equation}
Thus, the interface width is\cite{Bru}
\begin{equation}
w(\epsilon,t)\sim\left\lbrace \begin{matrix} t^{\beta}, & ~~ t\ll \ep^z \\
t^{\beta_*} \ep^H, &~~ t\gg \ep^z \end{matrix}\right.,
\end{equation} 
where $\beta_* = (\alpha-H)/z$.
The standard self-affine Family-Vicsek scaling is recovered if $\alpha=H$. 

In the present work, we report the dynamical scaling analysis of CdTe films grown on glass substrates covered by fluorine doped tin oxide. In addition to the $H$ and $\beta$ exponents previously reported,\cite{Igor} we calculated  the dynamic and roughness exponents, $z$ and $\alpha$.  With the complete set of exponents it was possible to identify new properties in this system, like anomalous scaling and super-roughness, which were not reported previously.  The paper is organized as follows. In section \ref{methods} the methodology is described. In section \ref{results} the results are presented and discussed and, finally, some conclusions are drawn in section \ref{conclusions}.

\section{\label{methods}Methods}
\begin{figure}[ht!]
\begin{center}
\includegraphics[width=8.5cm]{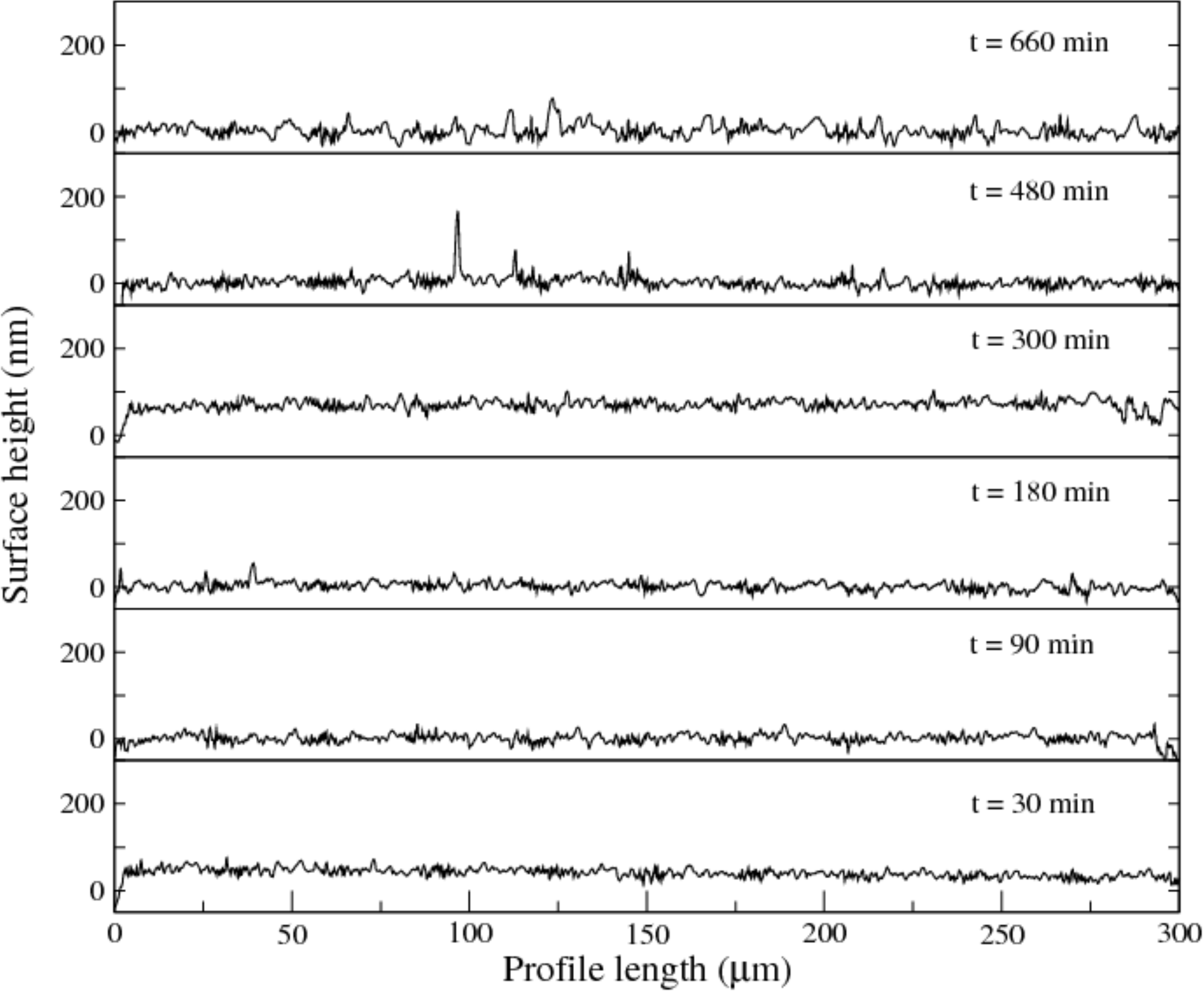} \\ (a) \\
\includegraphics[width= 8.5cm]{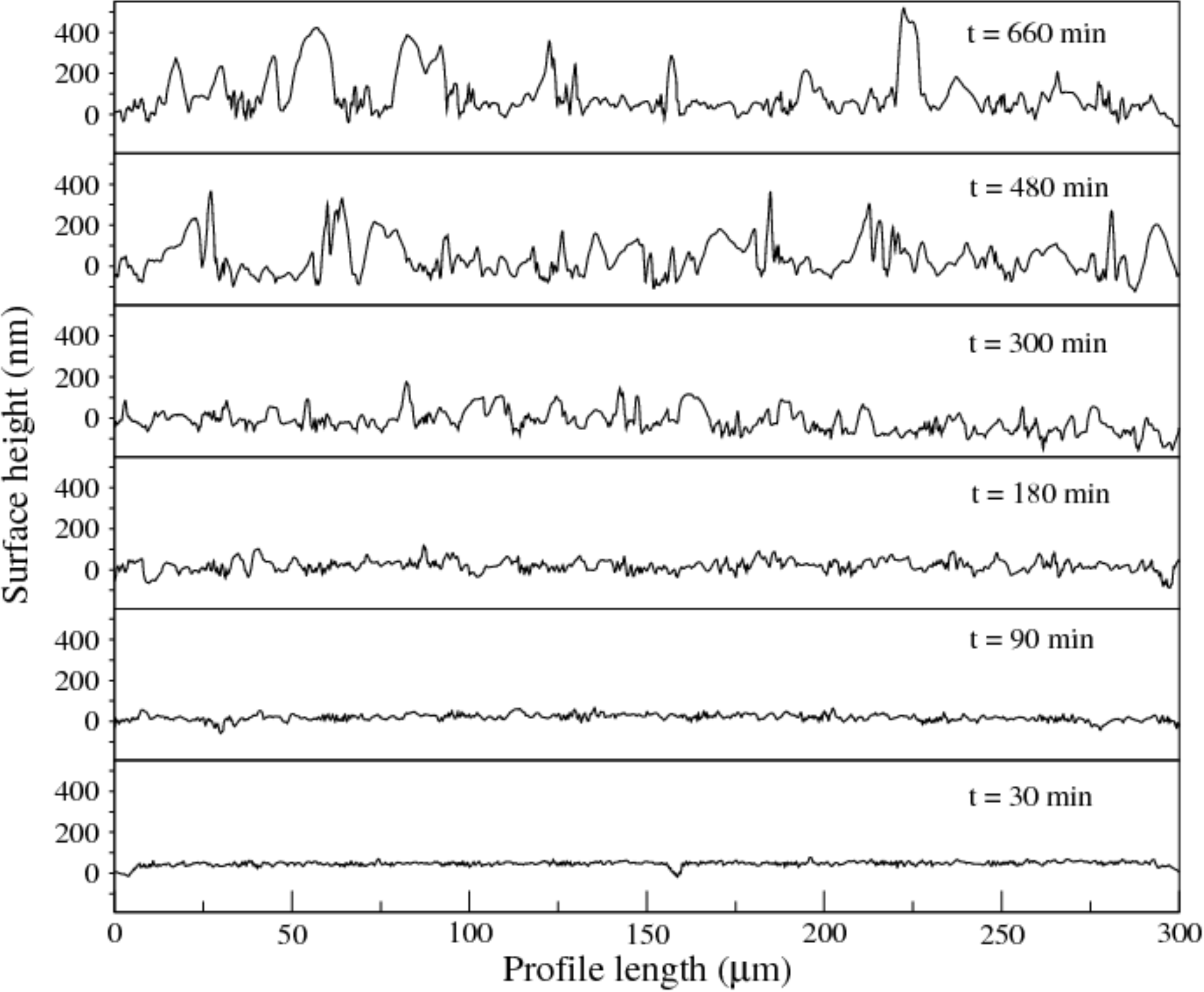} \\ (b)
\caption{\label{fig:profiles} Typical time sequence for scanned profiles of CdTe films grown at substrate temperatures (a) $T=150$ $^\circ$C and (b) $T=300$ $^\circ$C.}
\end{center}
\end{figure}

CdTe films were grown on glass substrates covered by fluorine doped tin oxide by Hot Wall Epitaxy (HWE).  
CdTe films were produced with a growth rate of 1.4 \AA/s and growth times varying from $t =$ 30 to 660 min at substrate temperatures varying from $T=$ 150 to 300 $^\circ$C. At least 20 surface profiles with length of 300 $\mu$m were measured for each sample, using a stylus profiler (XP1-AMBIOS) with vertical resolution better than 10 \AA~and lateral resolution of 10 nm. Details of the experimental setup can be found elsewhere.\cite{Leal} 

In figure \ref{fig:profiles}, we show time sequences of scanned profiles of CdTe films illustrating the increasing of the interface width with substrate temperature previously reported.\cite{Igor} One can clearly beheld in figure \ref{fig:profiles} that growth instabilities characterized by  sharp peaks emerge after a certain deposition time and are enhanced during the interface growth. Moreover, the instabilities are more pronounced  at higher temperature.

 Growth and Hurst exponents for the samples used in this work were reported previously,\cite{Igor} but the global roughness analysis was missing.  Indeed, several experimental limitations prevent the production of samples with very long deposition times and, consequently, the usual relation $w\sim L^\alpha$ for $t\rightarrow\infty$ cannot be used.\cite{Barabasi}  Alternatively,  the global roughness exponent $\alpha$  can be found  applying the scaling relation $\alpha=\beta z$, since growth and dynamical exponents can be determined for short deposition times using the scaling laws  $w\sim t^\beta$ and $\xi\sim t^{1/z}$, respectively.

The correlation length was determined using the two-point correlation function $\Gamma(\ep,t)$. For the sake of reproducibility, we used two distinct definitions to calculate  $\Gamma$.  The first one is the standard two point correlation function 
\begin{equation}
\label{eq:corr2}
\Gamma(\ep,t) = \left\langle\overline{\tilde{h}(x+\ep , t) \tilde{h}(x,t)} \right\rangle,
\end{equation}
where $\tilde{h}$ is the detrended profile.\cite{detrend} The second one is the probability of the height difference between two sites separated by a distance $\ep$ being lower than a fixed value $m$, {\it i e.}, 
\begin{equation}
\Gamma(\ep ,t) = \left\langle\overline{\Pr \left ( |\tilde{h}(x+\ep , t) - \tilde{h}(x , t)| \le m \right)} \right\rangle.
\label{eq:corr1}
\end{equation}
Where $m$ is much shorter than the global interface width and much larger than the profile height resolution. The correlation function given by equation (\ref{eq:corr1}) is analogous to the two-particle correlation function commonly applied to fractal aggregates.\cite{Meakin} Definitions (\ref{eq:corr2}) and (\ref{eq:corr1}) provide proportional correlation lengths and,  consequently, the same dynamical exponent. Brackets and bars follow the same notation defined in equation (\ref{eq:eq1}). For the experimental data $m=0.1|h_{max} - h_{min}|$ was used, where $h_{max}$ and $h_{min}$ are the maximum and the minimum heights in the detrended profile.

In both definitions, $\Gamma(\ep)$ was fitted by a two-exponential decay function
\begin{equation}
\label{eq:nonlin}
\Gamma(\ep,t) = \Gamma_0 + A_1\exp\left(-\frac{\ep}{\xi_1}\right) + A_2 \exp\left(-\frac{\ep}{\xi_2}\right)
\end{equation}
where the fit parameters $\Gamma_0$, $\xi_1$ and $\xi_2$ are time functions. {The two-exponential decay is just a generalization of the standard exponential decay commonly used for correlation function fits. Our alternative choice is due to equation (\ref{eq:nonlin}) fits very well all experimental data, while single exponential decay returns unsatisfactory fits for some samples.} So, the correlation length $\xi(t)$ is determined by 
\begin{equation} \int_0^\xi [\Gamma(\ep,t) - \Gamma_0] d\ep = 
f \int_0^\infty [\Gamma(\ep) - \Gamma_0] d\ep,
\end{equation}
where $0\ll f<1$ is a fraction determining a characteristic decay. In the present work, we choose $f=0.1$.

\begin{figure}[!ht]
\begin{center}
\includegraphics[width=8.5cm,height=!,clip=true]{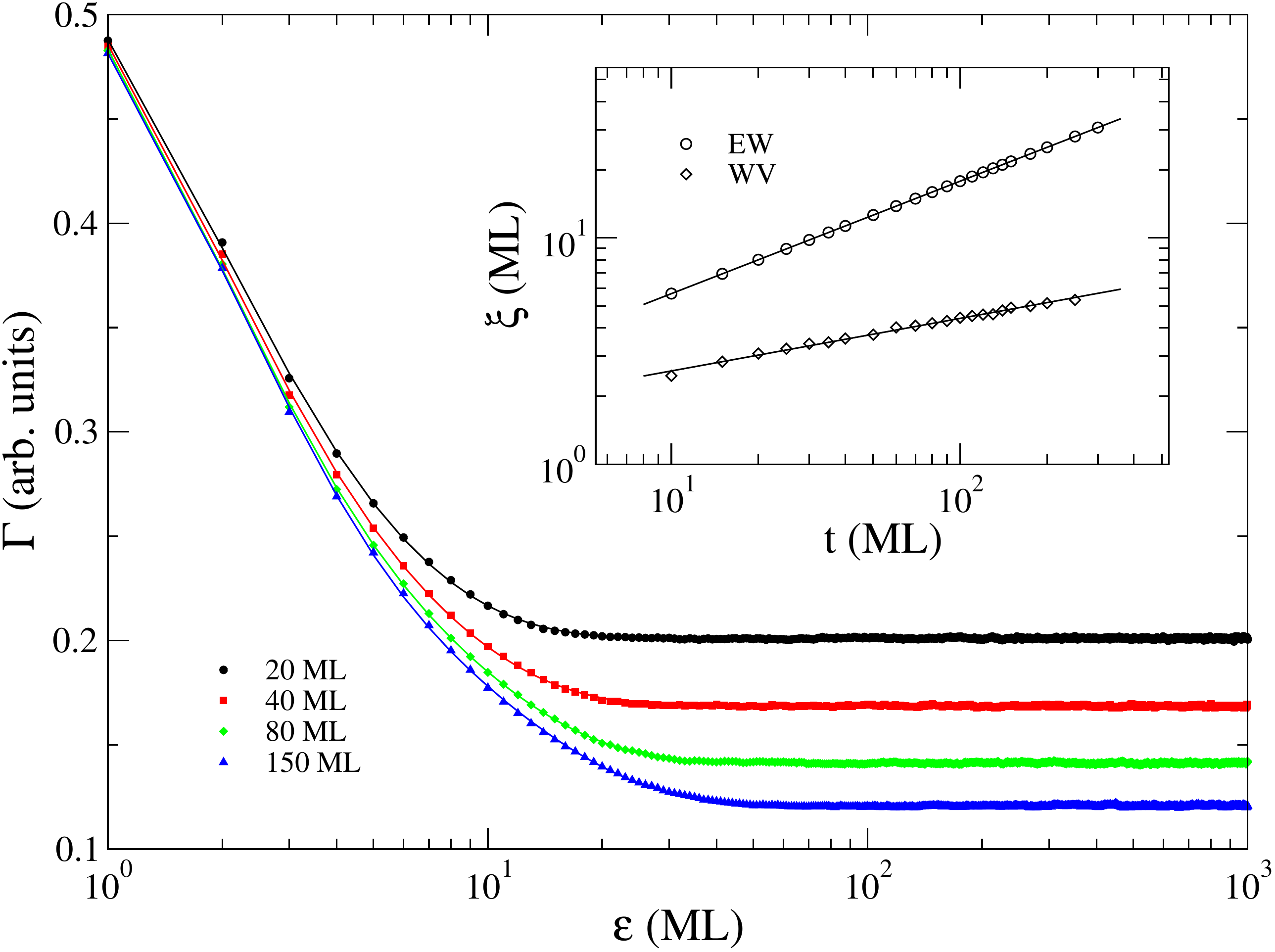} 
\caption{\label{fig:corrEW} (Color on-line) Correlation function (\ref{eq:corr1}) for the EW model at different deposition times. Symbols are simulational data and solid lines are non-linear fits using Eq. (\ref{eq:nonlin}). Inset shows the correlation length as function of time for the EW and WV models. The slopes $1/z = 0.496$ and $1/z = 0.236$ for EW and WV models, respectively, are in very good agreement with the 1/2 and 1/4 expected values.\cite{Barabasi} Here, ML means monolayers.}
\end{center}
\end{figure}

In figure \ref{fig:corrEW}, the last method was applied to two distinct models in order to verify the validity of equation (\ref{eq:corr1}). In the first one, the Edwards-Wilkinson (EW) model,\cite{Barabasi} particles are deposited on random sites of a linear chain and relax to the lowest neighboring position if this minimizes the particle height. In the second one, the Wolf-Villain (WV) model,\cite{Wolf} the particles are deposited at random and move to the nearest neighbor site that maximizes the number of bonds. As one can see, correct exponents were obtained for both models (see figure caption). The method was also successfully applied to other well-established models.

\section{\label{results} Results}

\begin{figure}[!ht]
\begin{center}
\includegraphics[width=8.5cm,height=!,clip=true]{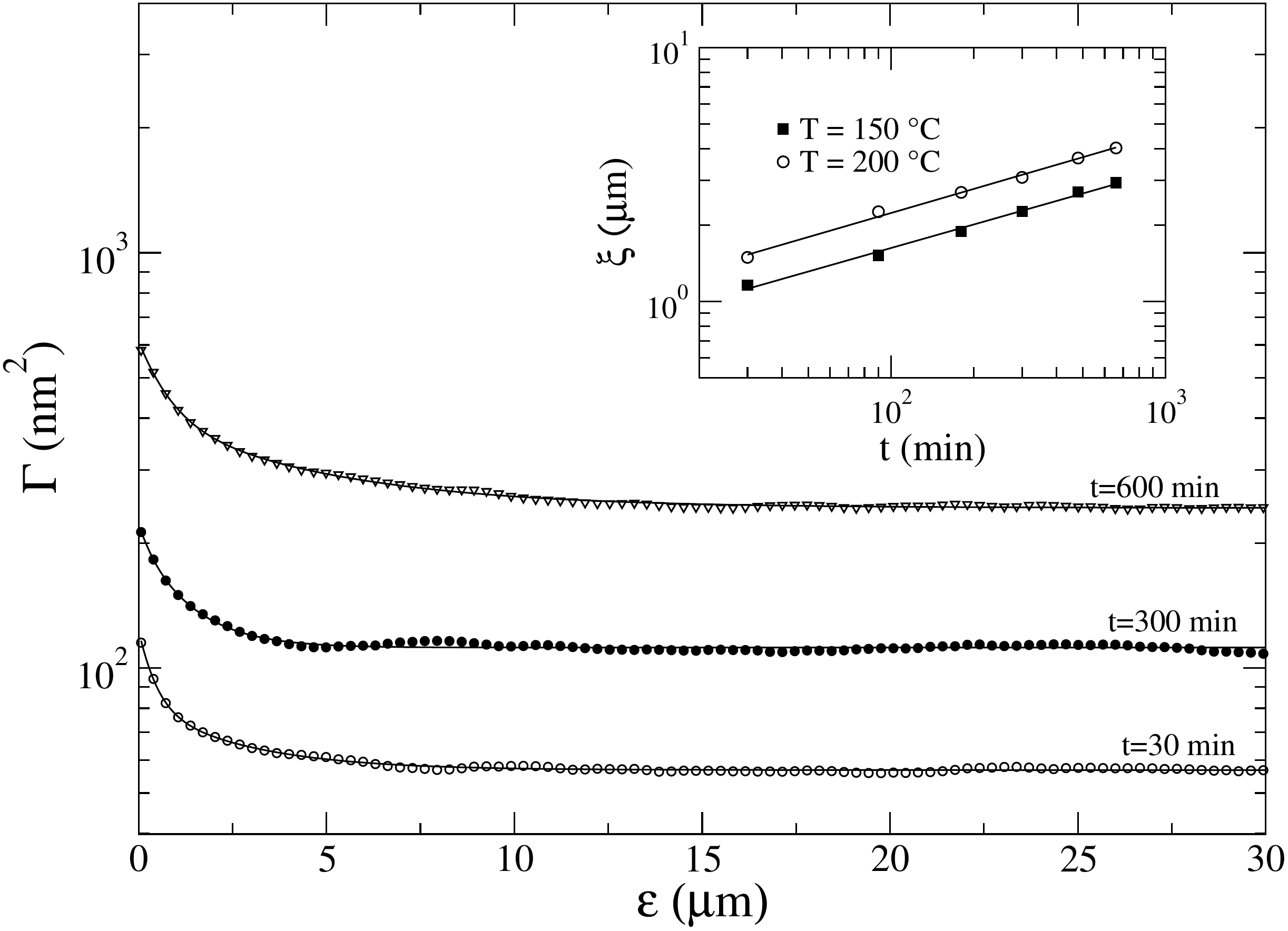}
\caption{\label{fig:correxp_novo}Correlation functions for the polycrystalline CdTe films grown at $T=200~^\circ$C using definition (\ref{eq:corr2}). Symbols are experimental data averaged over at least 20 surface profiles at each growth time and lines are nonlinear fits given by equation (\ref{eq:nonlin}). In the inset, the time dependence of the correlation length is shown for two distinct temperatures. }
\end{center}

\end{figure}

\begin{figure}[!ht]
\begin{center}
\includegraphics[angle=270,width=8.5cm,clip=true]{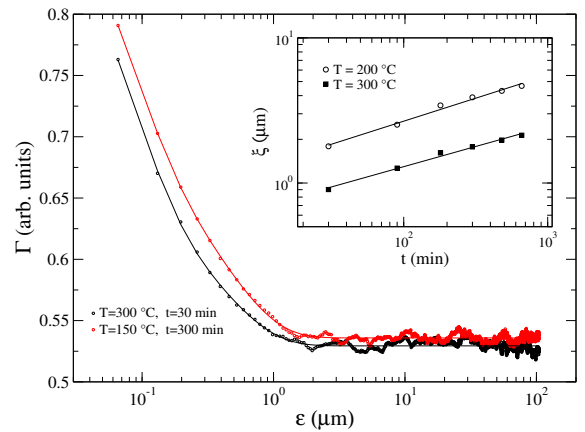}
\caption{\label{fig:correxp}(Color on-line) Correlation functions for the polycrystalline CdTe films using definition (\ref{eq:corr1}). Symbols are experimental data and lines nonlinear fits. In the inset, the time dependence of correlation length is shown for two distinct temperatures.}
\end{center}
\end{figure}

Correlation functions using equation (\ref{eq:corr2}) for $T = 200 ^\circ$C at distinct times are shown in figure \ref{fig:correxp_novo} and two representative curves using definition (\ref{eq:corr1}) are shown in figure \ref{fig:correxp}. In both figures, the insets show the correlation length as a function of time for distinct temperatures, demonstrating the validity of the power law $\xi\sim t^{1/z}$.


\begin{figure}[hbt]
\begin{center}
\includegraphics[width=8.5cm,height=!,clip=true]{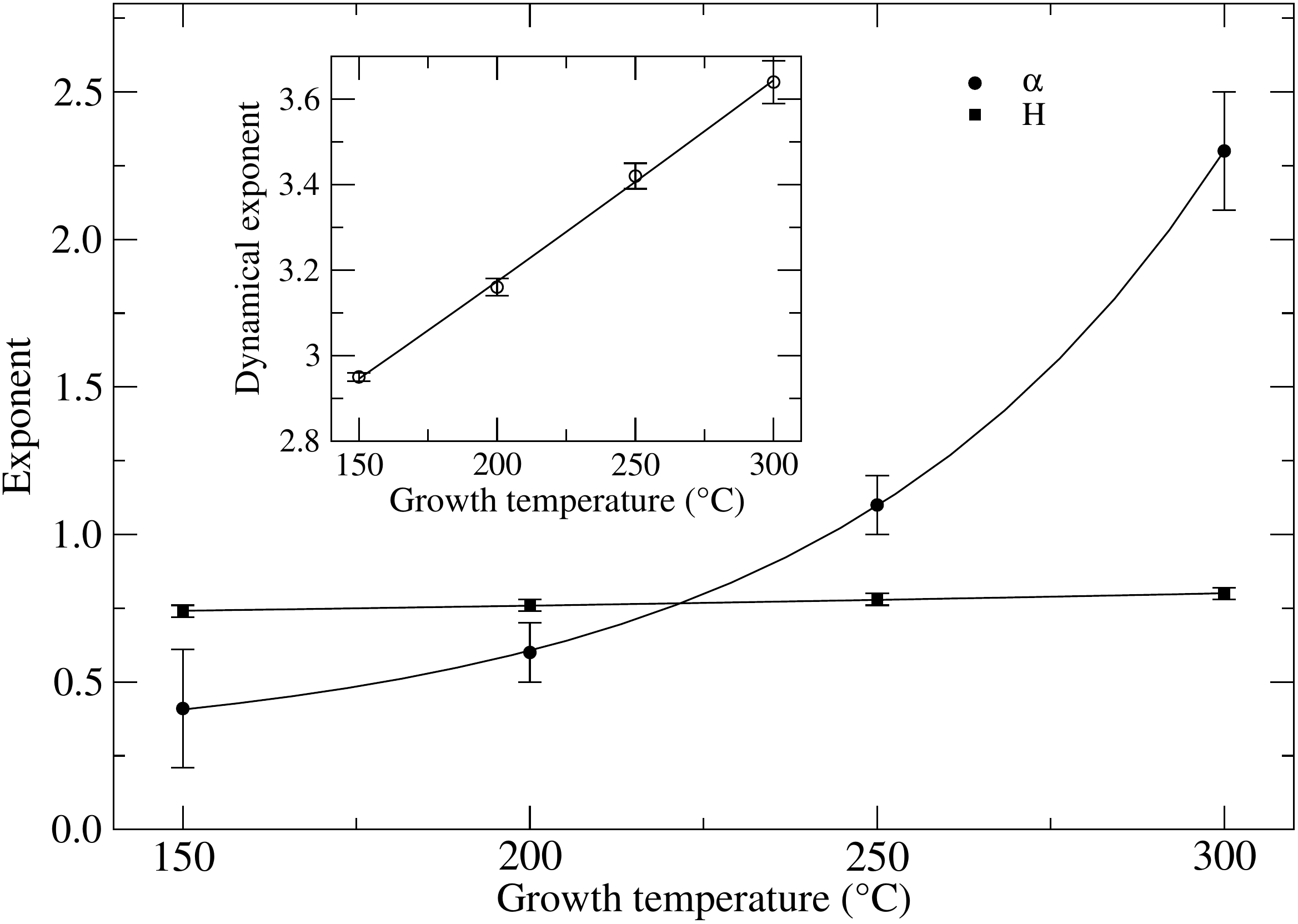}
\caption{\label{fig:exponents} Hurst and roughness exponents for CdTe films as functions of the substrate temperature. The inset shows the corresponding curves for the dynamical exponent. The values of the Hurst exponent were taken from Ferreira \textit{et al.} \cite{Igor} for $t=5$h. Solid lines are nonlinear fits to guide the eyes.}
\end{center}
\end{figure}

The dynamical exponent $z$ as a function of temperature is shown in the inset of figure \ref{fig:exponents},  in which the error bars are the standard deviations of the exponents obtained for at least 4 groups of 5 profiles at each temperature and growth time. The exponent increases  with the substrate temperature, implying in a slower spreading of the correlations along the surface and consequently an increase of the interface width saturation time as temperature increases. The global roughness exponent was obtained from the scaling relation $\alpha = \beta z$ using the values of the growth exponents $\beta$  taken from a previous work.\cite{Igor} The results are shown in figure \ref{fig:exponents}. Notice that the slow increase of the local roughness exponent $H$ with temperature contrasts with the fast increase observed for $\alpha$. Moreover, the system exhibits the so called \textit{intrinsically anomalous} roughening,\cite{Lopez2} for which the local roughness exponent $H<1$ is actually an independent exponent and $\alpha$ may take values larger or smaller than 1. Indeed, theoretical arguments claim that symmetries and conservation laws restrict the emergence of intrinsically anomalous roughening in growth process controlled by nonlocal effects \cite{Lopez2} such as quenched disorder and shadowing effects. It is very difficult to point out the nonlocal effects which could be ruling the growth dynamics of the samples studied here, since they probably depend on details of the growth process. Theoretical studies and new experiments, including the investigation of other materials and substrates, are necessary to elucidate this point.

\begin{figure}[hbt]
\begin{center}
\includegraphics[width=8.5cm,height=!,clip=true]{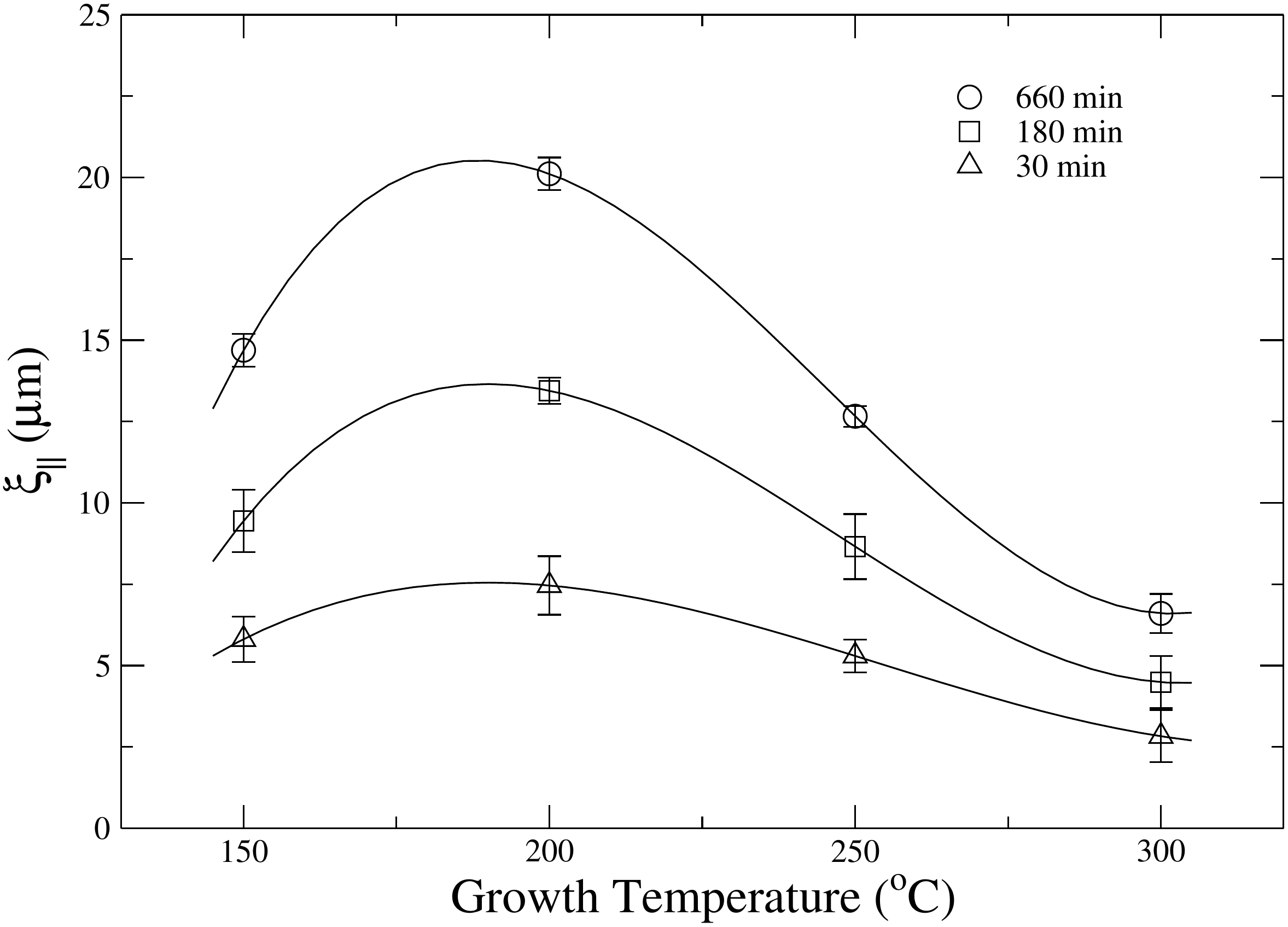}
\caption{\label{fig:correl_temp} Correlation length as a function of temperature for distinct times.}
\end{center}
\end{figure}

In figure \ref{fig:correl_temp}, the lateral correlation length $\xi_{\parallel}$ calculated using equation (\ref{eq:corr2}) is shown as a function of temperature for a fixed time. For low temperatures, the correlation length $\xi_{\parallel}$ increases with temperature as it is expected since the larger the diffusion rates the larger the correlation spreading. However, the opposite is observed at higher temperatures, which reflects the onset of peaks, i.e., the transition to the regime ruled by instabilities.  The microscopic origin of this behavior is not clear.  Exponents obtained from correlation functions (\ref{eq:corr2}) and (\ref{eq:corr1}) diverge at less that 10 \% for any temperature, assuring the reproducibility of the results.

\section{\label{conclusions} Conclusions}

CdTe films grown on glass substrates covered by fluorine doped tin oxide by Hot Wall Epitaxy (HWE) were studied using the interface dynamical scaling theory. We determined the dynamical exponent which revealed an intrinsically anomalous scaling characterized by a global roughness exponent $\alpha$ distinct from the local one, the Hurst exponent $H$.
In addition to the control of surface width formerly reported,\cite{Igor} we showed that the scaling exponents and, consequently, the growth dynamics in the surface, can be modified by varying the control parameters, in particular the substrate temperature. The transition from $\alpha<1$ to $\alpha>1$ has a particular importance which can be observed in figure \ref{fig:profiles}. For $T=150^\circ$C, when $\alpha<1$, the interface fluctuations grow approximately uniformly along the surface, while for $T=300^\circ$C, when $\alpha>1$, the surface is ruled by instabilities.

As discussed by Lop\'ez \textit{et al},\cite{Lopez2} symmetries and conservation laws restrict the emergence of the intrinsically anomalous roughening in local growth models. Indeed, disorder and/or nonlocal effects are required for intrinsic anomalous roughening. Our experiments are consistent with this hypothesis if we suppose that the original substrate introduces a sort of nonlocal disorder, which amplifies the instabilities along the surface growth. It was observed in simulations of epitaxial growth with a quenched disorder in the step barriers,\cite{Elsholz} that the growth exponent $\beta$ is an increasing function of temperature. In addition, the shadowing effects are probably strong in HWE technique due to the collimation vapor directed to the film.\cite{Leal} In summary,  our results are in accordance with the conjecture that the emergence of the intrinsically anomalous roughening in local growth process requires the break of symmetries and/or conservation laws. Following the simulational results by Elsholz et al. \cite{Elsholz}, we propose that the nonuniformity  of the original substrate introduces a quenched disorder as the cause of symmetry break, which is enhanced by a nonlocal shadowing effects came from the HWE deposition method. Theoretical studies and new experiments using other materials and substrates are necessary to elucidate this point.

\begin{acknowledgments}
This work was supported by Brazilian agencies FAPEMIG, CNPq, and CAPES. 
\end{acknowledgments}


\begin{thebibliography}{99}
\bibitem{detrend} $\tilde{h}(x) = h(x)-h_{fit}$, where $h_{fit}$ is the linear regression of the original profile $h(x)$.
\end{thebibliography}
\end{document}